\documentclass[twocolumn,floatfix,prl]{revtex4-2}%
\usepackage[dvipdfmx]{graphicx}%
\usepackage{amsmath}%
\usepackage{amsfonts}%
\usepackage{amssymb}
\usepackage{bm}
\usepackage{color}
\usepackage{ulem}

\def\e{{\epsilon}}
\def\k{{ {\bm k} }}
\def\p{{ {\bm p} }}
\def\q{{ {\bm q} }}
\def\Q{{ {\bm Q} }}

\def\0{{ {\bm 0} }}
\def\w{{\omega}}
\def\a{{\alpha}}

\allowdisplaybreaks[4]

\begin{document}
\title{
Three-dimensional bond-order instability in infinite-layer
nickelates due to nonlocal quantum interference}
\author{
Seiichiro Onari and Hiroshi Kontani
}

\date{\today }

  \begin{abstract}
Recently discovered superconducting infinite-layer nickelates $R$NiO$_2$
   ($R$=Nd, La, Pr) attracts increasing attention as a similar
   system to cuprates. Both $R$NiO$_2$ and YBCO
   cuprates display the three-dimensional (3D) CDW with wave vector
   $\q\sim(2\pi/3,0,q_z)$, while $q_z$ is non-zero and incommensurate in the former system.
Here, we reveal that the characteristic CDW in $R$NiO$_2$ 
   can be \color{black} explained
   as the quantum interference between paramagnons,
   by focusing on the following characteristics of $R$NiO$_2$:
 (i) prominent three-dimensionality in the Fermi surface and (ii) large
   self-hole-doping ($\sim 14$\%).  This mechanism predicts the emergence of the
   $d_{x^2-y^2}$-wave bond order at a secondary 3D nesting vector
$\q^c\sim(2\pi/3,0,q^c_z)$ $(q^c_z=0.2\pi\sim 2\pi/3)$. 
  The obtained strong bond fluctuations lead to the
   non-Fermi liquid electronic states and superconducting states in
   nickelates. 
  \end{abstract}

\address{
 Department of Physics, Nagoya University,
Furo-cho, Nagoya 464-8602, Japan. 
}
 
%\pacs{74.70.Xa, 75.25.Dk, 74.20.Pq} 
%74.20.Pq Electronic structure calculations

\sloppy

\maketitle

%%%%%%%%%%%%%%%%%%
%Introduction
%%%%%%%%%%%%%%%%%%
The recent discovery of superconducting infinite-layer nickelates
with similar electronic structures and phase diagrams to cuprates has
stimulated much attention \cite{Ni-super1,Ni-super2,Ni-super3}.
In fact, infinite-layer nickelates $R$NiO$_2$ ($R$=Nd, La, Pr) have Ni-$3d^9$
configuration, which is the same as Cu-$3d^9$ in cuprates. 
A high superconducting transition temperature $T_{\rm SC}\lesssim30$K has been
reported in $R$NiO$_2$ \cite{Ni-super4}, which may be due to the
electron structure similar to cuprates.

However, there are many differences between $R$NiO$_2$ and cuprates.
The charge-transfer energy estimated by experiments and theories in
$R$NiO$_2$ is larger than cuprates
\cite{Ni-first-principle1,Ni-first-principle2,Ni-first-principle3,Ni-first-principle4,Ni-Sakakibara,Ni-first-principle5}.
Thus, $R$NiO$_2$ is close to the Mott-Hubbard regime, away from the charge-transfer
regime in Zaanen-Sawatzky-Allen classification
\cite{Zaanen-Sawatzky-Allen}.
Significant differences from cuprates
   are (i) prominent three-dimensionality in the Fermi surface (FS) and (ii) large self-hole-doping.
As for (i), the FS composed of Ni $d_{x^2-y^2}$ orbital in NdNiO$_2$ has three-dimensionality, while cuprates have
two-dimensional (2D) FS.
As for (ii), self-hole-doping ($p_{\rm self}\sim$0.14) for
Ni $d_{x^2-y^2}$ orbital in NdNiO$_2$
is induced since the FSs of the Nd orbitals
emerge \cite{Ni-first-principle2,Ni-first-principle3}.
 We define an effective
 hole-doping $p_{\rm eff}=x+p_{\rm self}$ in Nd$_{1-x}$Sr$_x$NiO$_2$ to
 make direct comparison with $p_{\rm eff}=x$ in cuprates without the
 self-hole-doping. 

 Very recently, a three-dimensional (3D) CDW at wave vector
 $\q=(q_x,0,q_z)$ $(q_x\sim2\pi/3,q_z\ne0)$ has been observed by RIXS measurements
 in $R_{1-x}$Sr$_x$NiO$_2$ \cite{Ni-CDW1,Ni-CDW2,Ni-CDW3,Ni-CDW4}.
The value of $q_x\sim2\pi/3$ corresponds to the period-three CDW in the
 $x$-direction. \color{black} The value of $q_x$ and the CDW transition
 temperature ($T_c^{\rm RIXS}$) observed by RIXS measurements decrease
 with $p_{\rm eff}$.
 Schematic $p_{\rm eff}$
 dependences of $T_c^{\rm RIXS}$ observed
 in nickelates and overdoped cuprates are shown in Supplementary
 Material (SM) A \cite{SM}. \color{black}
 $T_c^{\rm RIXS}\lesssim 400$K in nickelates is higher than $T_c^{\rm RIXS}\lesssim 200$K in cuprates. 
 In both systems, the CDW quantum critical
 point (QCP) attracts great attention since $T_{\rm SC}$ becomes the
 maximum around the CDW QCP. In nickelates, critical hole-doping is $p_c^{\rm eff}\sim
 0.25$ \cite{Ni-CDW1}, while $p_c^{\rm
 eff}\sim 0.18$ in YBCO cuprates \cite{Cu-CDW,Cu-CDW2}.
Near the CDW QCP, the non-Fermi-liquid transport
 phenomena have been observed
 \cite{Ni-Non-Fermi1,Ni-Non-Fermi2,Non-Fermi-Cu,Non-Fermi}.
 Interestingly, the pseudogap observed in Bi2212 for
 $p_{\rm eff}<0.19$ \cite{Bi2212-1,Bi2212-2} would originate from the
 CDW formation. In contrast, Ref. \cite{e-doped-cuprate} reports the
 absence of the CDW in electron-doped cuprates. \color{black}
 
 The electronic states in $R$NiO$_2$ have been actively studied by using the dynamical mean-field theory (DMFT) \cite{Ni-DFT-DMFT1,Ni-DFT-DMFT2}.
Beside the DMFT, the pair-density-wave
 \cite{Cu-PDW-theory1,Cu-PDW-theory2,Cu-PDW-theory3},  intertwined-order \cite{Cu-PDW-theory1,Davis-DHLee}, spin-nematic/vestigial-order
\cite{Fernandes,Fernandes-122,DHLee,QSi,Valenti,Fang,Fernandes-review}, and 
orbital/bond-order \cite{Kruger,PP,WKu,Kontani-PRL,Onari-SCVC,Onari-form,Yamakawa-PRX,Onari-B2g,Onari-smectic,Onari-Frontier,Adv-review,JP,Fanfarillo,Chubukov-FeSe,Chubukov-RG,Cu-CDW-VC1,Cu-CDW-VC2,Cu-CDW-VC3,Tsuchiizu-Ru}
scenarios have been proposed to explain the CDW.
According to Refs. \cite{Cu-CDW-VC1,Cu-CDW-VC2,Cu-CDW-VC3,Kontani-sLC}, the bond
 order and spin current order in cuprates can be explained by ``the paramagnon-interference mechanism''
 described by the non-local vertex corrections. \color{black} The similar theory has been applied to Fe-based
superconductors
\cite{Onari-SCVC,Onari-form,Yamakawa-PRX,Onari-B2g,Onari-smectic,Onari-Frontier},
 twisted bilayer graphene \cite{Onari-TBG} and kagome
 metal \cite{Tazai-kagome-CDW,Tazai-kagome-CLC}.
 
In this paper, we study the origin of the 3D CDW in $R$NiO$_2$ based
on the paramagnon-interference mechanism by using
the 3D density-wave (DW) equation.
We find that the 3D bond order with wave vector
$\bm{q}^c=
(2\pi/3,0,q^c_z)$ $(0.2\pi\lesssim q^c_z \lesssim 2\pi/3)$ emerges, which is consistent with the 3D CDW with
$q_z\sim0.6\pi$ and $q_z\lesssim0.54\pi$ observed in experiments
\cite{Ni-CDW3}. 
The obtained $p_{\rm eff}$ dependences of the bond order are
consistent with experiments \cite{Ni-CDW1}.
 The obtained strong bond fluctuations lead to the non-Fermi
liquid states and superconducting states \cite{Tazai-kagome-CDW}.

\begin{figure}[!htb]
\includegraphics[width=.99\linewidth]{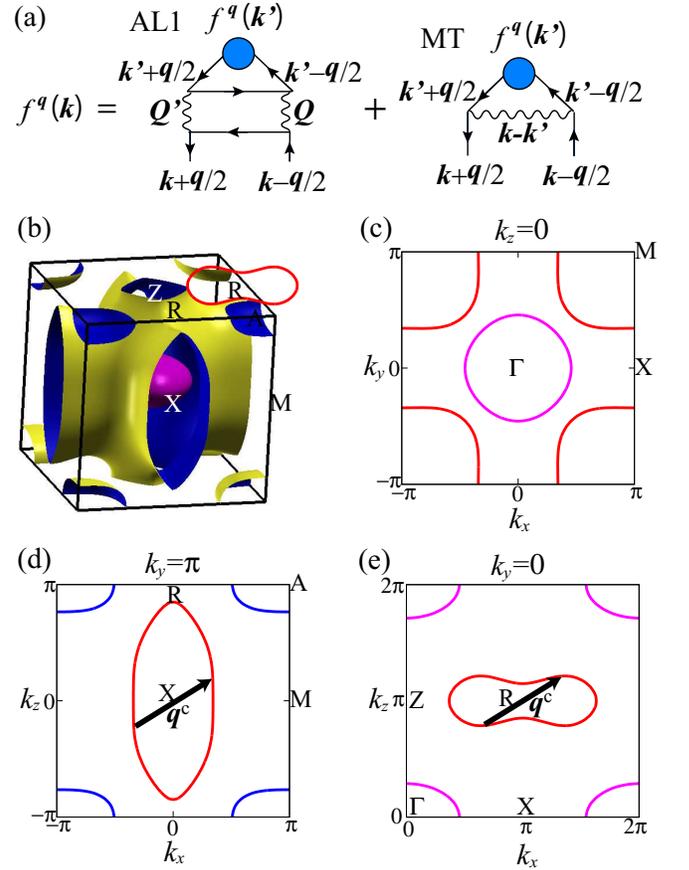}
\caption{
%(color online)
 (a) Feynman diagrams of the AL1 and MT terms in the DW Eq., where the wavy lines
 represent the spin fluctuations.
  (b) 3D FSs in the present NdNiO$_2$ model, where the red line
 represents FS around the R point shown in Fig. \ref{fig1} (e).
 (c) FSs on $k_z=0$ plane.
 (d) FSs on $k_y=\pi$ plane.
  (e) FSs on $k_y=0$ plane.
 Red, purple, and blue lines show FSs composed of orbitals 1, 2, and 3,
 respectively.
 Black arrows are secondary 3D nesting vector
 $\q^c\sim(2\pi/3,0,q^c_z)$ $(0.2\pi\lesssim q^c_z\lesssim 2\pi/3)$.
}
\label{fig1}
\end{figure}

The 3D bond order is derived by the paramagnon interference due to the
Aslamazov--Larkin (AL) terms in Fig.\ref{fig1}(a), which are non-local irreducible vertex corrections. In the paramagnon-interference mechanism, the charge channel order with $\q=\Q-\Q'$ originates from the interference between the spin fluctuation with $\q=\Q$ and that with $\q=\Q'$. 
This mechanism is not derived from the mean-field-like approximation, such as the random-phase-approximation (RPA) and the fluctuation-exchange approximation without the vertex corrections. The present paramagnon-interference mechanism would be a key concept toward a unified understanding of the CDW orders in nickelates and cuprates.
\color{black}

%%%%%%%%%%%%%%%%%%%%%%%%%

%%%%%%%%%%%%%%%%%%%%%%%%%%%
% Model and Hamiltonian
%%%%%%%%%%%%%%%%%%%%%%%%%%%
We analyze the following 3D
three-orbital Hubbard model for NdNiO$_2$, where $d_{x^2-y^2}$ orbital of Ni,
$d_{z^2}$ and $d_{xy}$ orbitals of Nd are taken into account:
$H=H^0+H^U$, where $H^0$ is the tight-binding model for NdNiO$_2$ based on
Ref. \cite{Ni-first-principle4}. We introduce the next-nearest-interlayer hopping $t_{z2}$ of Ni $d_{x^2-y^2}$ orbital in order to reproduce
the parallel FSs around the M point obtained by other
first-principles calculations
\cite{Ni-Sakakibara,Ni-first-principle5}.
Details of the model are explained in SM A \cite{SM}.
The orbitals $1$, $2$, and $3$ denote Ni $d_{x^2-y^2}$ orbital,
Nd $d_{z^2}$ orbital, and Nd $d_{xy}$ orbital, respectively.
 $H_U$ is the Coulomb interaction, where the Coulomb
interaction of only
orbital 1 is taken into account since the results including the Coulomb
interactions of the three orbitals are almost the same. \color{black}

Figures \ref{fig1}(b) and S1(a) in SM A \cite{SM} show
3D FSs and band dispersion in this model $(x=0)$, where the number of electrons
is 1. Since the FSs composed of orbitals 2
and 3 appear, the number of electrons for the Ni $d_{x^2-y^2}$ orbital is about $0.86$
(self-hole-doping $p_{\rm self}=0.14$),
which is consistent with the self-hole-doping reported in
Ref. \cite{Ni-first-principle2,Ni-first-principle3}.
In Nd$_{1-x}$Sr$_x$NiO$_2$, the value of $p_{\rm self}$ decreases with doping
 $x$, and $p_{\rm self}\sim$0.09 for $x=0.15$.
 
Here, we discuss the FS of the Ni $d_{x^2-y^2}$ orbital, which is important to realize the CDW.
Figure \ref{fig1}(c) shows the FSs in $k_z=0$ plane, where the FS
composed of Ni $d_{x^2-y^2}$
orbital is similar to the FS of YBCO cuprates.
Figure \ref{fig1}(d) shows the FSs and 3D nesting
$\q^c\sim(2\pi/3,0,q^c_z)$ around X point.
The spin susceptibility at $\q^c$ is smaller than that at primary
nesting $\Q_s$.
%This secondary nesting between the parallel FSs gives a broad peak
%structure of the
%irreducible susceptibility $\hat{\chi}^0(q)$ along the $q_z$ direction
%as shown in Fig. S1 in SM A \cite{SM}, while its height is too small to induce the spin-density-wave order.
The nesting $\q^c$ at the FS around R point shown in
Fig. \ref{fig1}(e), which is absent in cuprates, assists the
secondary nesting.
The small instability by the secondary nesting causes the bond order at
$\q=\q^c$ $(0.2\pi\lesssim q^c_z\lesssim 2\pi/3)$
with the aid of the AL vertex correction in the
present theory as shown later.

%%%%%%%%%%%%%%%%%%%%%%%%%%%%%%%%%
%%%%%%%%%%%%%%%%%%%%%%%%%%%%%%%%%
Before discussing the CDW order, 
we calculate the spin susceptibility for orbital 1 ${\chi}^{s}(q)$ because we discuss the spin-fluctuation-driven CDW mechanism.
We obtain ${\chi}^{s}(q)$
for $q=(\q,\w_m=2m\pi T)$ based on the RPA,
which is introduced in SM A \cite{SM}.
${\chi}^{s}(q)\propto(1-\a_{s})^{-1}$, where $\a_{s}$ is the spin
Stoner factor. $\a_{s}=1$ corresponds to spin-ordered
state.
 Hereafter, we fix $T=60$meV$\sim 700$K using $k_B=1$ unless
otherwise noted. \color{black}

\begin{figure}[!htb]
\includegraphics[width=.99\linewidth]{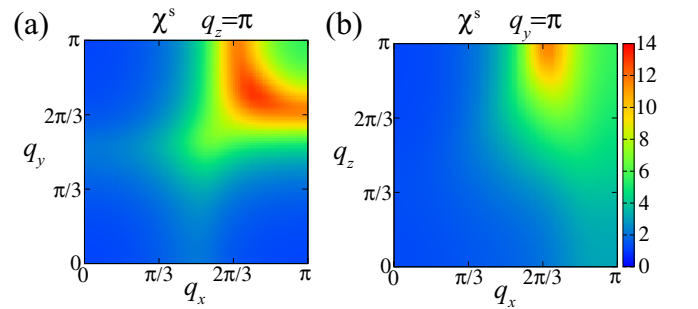}
 \caption{
(a) $\q$ dependences of 
 $\chi^{s}(\q,0)$ given by the RPA on $q_z=\pi$ plane, and
(b) that on $q_y=\pi$ plane in NdNiO$_2$.
}
\label{fig2}
\end{figure}

Figures \ref{fig2}(a) and (b) show the obtained spin susceptibility
for orbital 1
$\chi^{s}(\q,0)$, which has rather broad peak around $\Q_s=(2\pi/3,2\pi/3,\pi)$.
The value of $\chi^{s}(\q,0)$ becomes small away from the
$q_z=\pi$ plane, and other orbital components of spin
susceptibility are very small.
In the present study, we set the Coulomb interaction
$U_1=1.43$eV for the orbital 1, which is smaller than $U_1=3$-$4$eV obtained by the
first-principles calculations
\cite{Ni-Sakakibara,Ni-first-principle2} since
the self-energy is not included in the present study.
Approximately, the self-energy renormalizes the value of $U_1$ to $U_1/z$, where $z$ $(< 1)$ is the renormalization factor.
In order
to decide the value of $U_1$, we set the value of $\a_s=0.95$ at $x=0$ ($p_{\rm eff}=0.14$). 
The value of $\alpha_s$ decreases with $p_{\rm
eff}$. At $x=0.15$ ($p_{\rm eff}=0.24$),
 moderate spin fluctuations, $\alpha_s=0.91$, are obtained. \color{black}
These moderate spin fluctuations are consistent with the experiment, where 
$1/T_1T$ moderately increases for $T\lesssim 100$K \cite{Ni-NMR}.

%%%%%%%%%%%%%%%%%%%%%%%%%%%%%%%%%
%%%%%%%%%%%%%%%%%%%%%%%%%%%%%%%%%

Next, we analyze the CDW state in NdNiO$_2$ based on the
charge-channel DW equation \cite{Onari-form,Onari-B2g,Kontani-sLC}.
A rigorous formalism of the DW
equation has been constructed based on the Luttinger--Ward theory in
Ref. \cite{Tazai-LW}. The solution of the DW equation gives the minimum
of the grand potential in the Luttinger--Ward theory, and therefore it is
thermodynamically stable.
The optimized non-local form factor
${f}^\q(k)$ for orbital 1, which describes the DW order parameter, is derived from the following
 linearized DW equation:
\begin{eqnarray}
&& \lambda_\q {f}^\q(k)= \frac{T}{N}
\sum_{k'} {I}^{\bm{q}}(k,k'){g}^{\bm{q}}(k'){f}^\q(k'),
\label{eqn:linearized}    
\end{eqnarray}
$\lambda_\q$ is the eigenvalue of the form factor ${f}^\q(k)$, $g^{\bm{q}}(k)\equiv
-G\left(k+\frac{\bm{q}}{2}\right)G(k-\frac{\q}{2})$,
${I}^{\bm{q}}(k,k')$ is the four-point vertex, and $k=[\k,\epsilon_n=(2n+1)\pi T]$.
The charge-channel DW with wave vector $\q$ is established when
the largest $\lambda_\q=1$.
The DW susceptibility is proportional to
 $(1-\lambda_\q)^{-1}$ \cite{Tazai-LW}. Therefore, $\lambda_\q$ represents the strength
 of the DW instability.
In the DW Eq. (\ref{eqn:linearized}),
the Maki--Thompson (MT) terms and AL terms are
included in the four-point vertex, as we explain in SM A \cite{SM}.
Notably, the bond
order solutions in the square lattice and the anisotropic triangular
lattice Hubbard models \cite{Tazai-swave} obtained by the DW Eq. have been verified by the
renormalization group (RG) methods \cite{Chubukov-RG,Cu-CDW-VC3,Tsuchiizu-Ru}, where the higher-order
vertex corrections are generated in a systematic and unbiased manner.
\color{black}
When $\chi^s(\q)$ is large,
the AL term in Fig. \ref{fig1}(a) is strongly enhanced in proportion to $\sum_p\chi^s(\bm{p}+\q^c/2)\chi^s(\bm{p}-\q^c/2)$, where
$\Q=\bm{p}+\q^c/2$ and $\Q'=\bm{p}-\q^c/2$ close to the paramagnon wave
vector $\Q_s$ give dominant contribution.
 \cite{Onari-form,Onari-B2g}. \color{black} As a result, the interference mechanism
 causes the charge-channel DW order at $\q^c\approx\Q-\Q'$. The MT term is also important for the $\k$ dependence of ${f}^\q(k)$.

\begin{figure}[!htb]
\includegraphics[width=.99\linewidth]{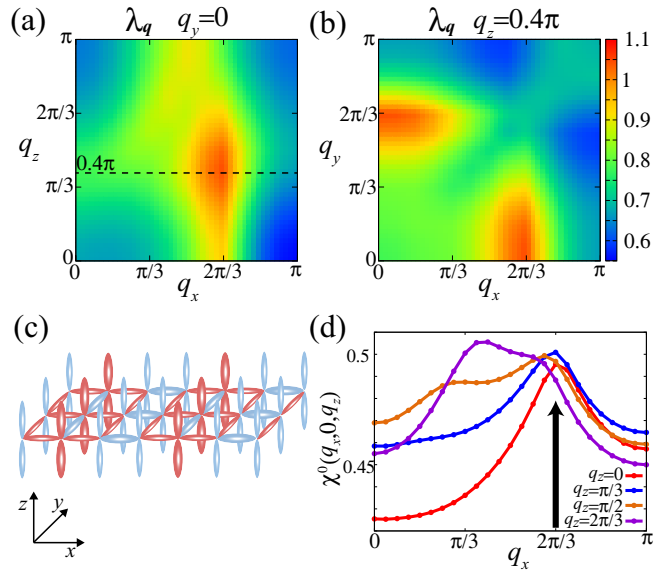}
 \caption{
(a) Obtained $\q$ dependence of $\lambda_{\q}$ on $q_y=0$ plane, and (b)
 that on $q_z=0.4\pi$ plane.
 (c) Schematic picture of 3D bond order, where the red and
 blue bonds denote increased and decreased hoppings. In addition, this
 bond order also has a modulation along the $z$-direction.
 (d)$q_x$ dependence of $\chi^0(q_x,0,q_z)$ for each $q_z$ at $x=0$. 
}
\label{fig3}
\end{figure}

 Figures \ref{fig3}(a) and (b) show the $\q$ dependence of the obtained
$\lambda_\q$, which peaks at the 3D secondary nesting
vector $\q^c$. 
From the Fourier transformation of ${f}^{\q^c}(\k)$ shown in SM B
\cite{SM}, the obtained order is identified as the $d_{x^2-y^2}$-wave
 bond order. The bond order has a period-three modulation in the
  $x$-direction \color{black} as shown in Fig. \ref{fig3}(c) due to
the $q_x$ component of $\q^c$ $(q^c_x\sim2\pi/3)$.
In the bond-ordered state, the hopping integrals are modulated. The obtained period-three bond order is consistent with the
experiments \cite{Ni-CDW1,Ni-CDW2,Ni-CDW3,Ni-CDW4}.
Since $\q^c$ has $q_z$ component
$q^c_z$, the bond order also has a modulation along
the $z$-direction, which is consistent with
the 3D CDW with 
$q_z\sim0.6\pi$ and $q_z\lesssim0.54\pi$ observed in Ref. 
\cite{Ni-CDW3}. 

%%%%%%%%%%%%%%%%%%%%%%%%%%%%%%%%%
%%%%%%%%%%%%%%%%%%%%%%%%%%%%%%%%%

In the following, we explain the decisive role of the non-locality of
the irreducible four-point vertex on the 3D bond
order
\cite{Onari-SCVC,Onari-form,Yamakawa-PRX,Onari-B2g,Onari-smectic,Onari-Frontier,Adv-review,Kontani-sLC}.
As discussed in Ref. \cite{Kontani-sLC}, in the presence of moderate spin fluctuations,
the AL terms give strong attraction between the Fermi momenta $\k$ and $\k'=\pm \k$ in the DW Eq. (\ref{eqn:linearized}), which leads to the relation $f^{\q^c}(\k)f^{\q^c}(\k')>0$ for $\k'=\pm \k$.
Thus, the AL terms strongly enhance the instability of various
even-parity $[f(\k)=f(-\k)]$ DW states.
In addition, the MT term in Fig. \ref{fig1}(a) gives moderate repulsion between the Fermi momenta $\k$
and $\k'=\k-\Q_s$, which leads to the relation
$f^{\q^c}(\k)f^{\q^c}(\k-\Q_s)<0$ at
$\Q_s\sim(2\pi/3,2\pi/3,\pi)$.
The MT term favors the $d$-wave form factor with sign reversal, as shown
in Figs. S2(a) and (b) in SM B \cite{SM}.
Due to the cooperation between the attraction by the AL terms and the repulsion by the MT term, the $d_{x^2-y^2}$-wave bond order is naturally realized at high transition temperature.
We note that the
non-local and non-$s$-wave DW states cannot be 
obtained when locally approximated $\hat{I}^{\bm{q}}_{\rm
local}(\epsilon_n,\epsilon_{n'})=\sum_{\k,\k'}\hat{I}^{\bm{q}}(k,k')$ is
applied, even if the AL and MT terms are taken into account
\cite{Tazai-swave}.
% 
%\color{red} Thus, the obtained bond order is difficult to be explained by the
%locally approximated methods such as the DMFT,
%the dynamical vertex approximation\cite{DgammaA}, and dual fermion\cite{DF}. \color{black}

Here, we also explain the importance of the secondary 3D nesting with short wavelength
shown in Fig. \ref{fig1}(d) and (e).
This nesting assists the $\bm{q}^c$ bond order. 
$g^\q(k)$ in the DW Eq. (\ref{eqn:linearized}) becomes large when both $\k+\frac{\bm{q}}{2}$ and
$\k-\frac{\bm{q}}{2}$ locate on the FSs.
The existence of this nesting is well recognized in a 
 broad peak structure of irreducible susceptibilities
 $\chi^0(\q)(\propto\sum_{k}g^\q(k))$ for orbital 1 as shown in
 Fig. \ref{fig3}(d). This secondary nesting stabilizes the 3D CDW in NdNiO$_2$.

We stress that the $\q=\bm{0}$ DW orders have been realized in various systems such
as cuprates and Fe-based superconductors, and
they can be naturally explained by 
the paramagnon-interference mechanism. The $\q$ dependence of DW order
is sensitive to the structure of FSs. The 
three-dimensionality in FSs might be important to understand the absence
of $\q=\bm{0}$ order. Clarifying the presence or absence
of $\q=\bm{0}$ order in $R$NiO$_2$ is an important future problem.
 
\begin{figure}[!htb]
\includegraphics[width=.99\linewidth]{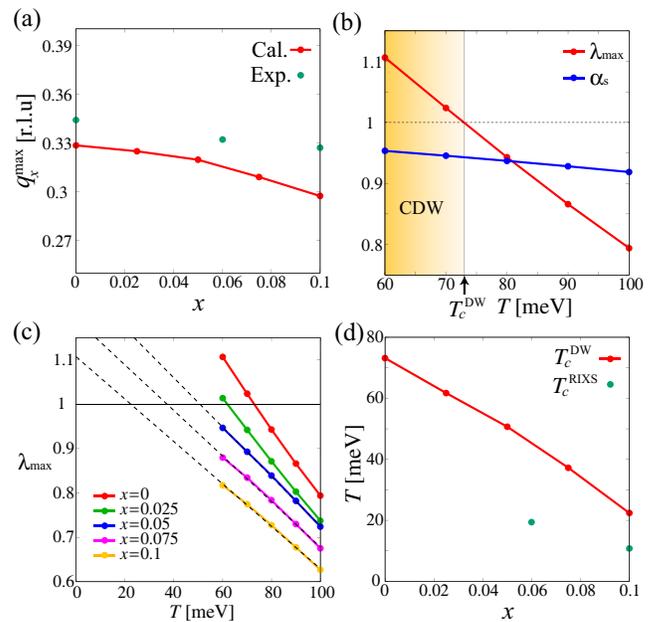}
\caption{
 (a) $x$ dependences of $q_x^{\rm max}$ in Nd$_{1-x}$Sr$_x$NiO$_2$.
 (b) $T$ dependences of $\lambda_{\rm max}$ and $\alpha_s$ for $x=0$ $(p_{\rm eff}=0.14)$.
(c) Obtained $T$ dependences of $\lambda_{\rm max}$ for
 $x=0,0.025,0.05,0.075,0.1$, where the dotted lines are fitted by the
 least squares.
 (d) $x$ dependences of $T_c^{\rm DW}$ and $T_c^{\rm RIXS}$.
 Lines are calculation results, while green dots represent experimental results in Ref. \cite{Ni-CDW1}.
}
\label{fig4}
\end{figure}

Here, we discuss the doping $x$ dependence of the CDW in Nd$_{1-x}$Sr$_x$NiO$_2$.
In this study, the hole-doping $x$ is introduced by the rigid-band
shift for all three bands. 
Figure \ref{fig4}(a) shows $x$ dependence of $q_x^{\rm
max}$. $q_x^{\rm
max}$ is defined as $q_x$ component of $\q$, where $\lambda_\q$ becomes maximum.
The value of $q_x^{\rm
max}$ decreases with hole-doping $x$ since the distance of the
two FSs around the M point, which is related to the 3D nesting $\q^c$, decreases with $x$. 
This $x$ dependence is consistent with the experimental
results \cite{Ni-CDW1}.

In order to derive the theoretical transition temperature $T_c^{\rm DW}$, we
calculate the $T$ dependences of the maximum value of $\lambda_{\q}$
$(=\lambda_{\rm max})$ and $\alpha_s$. 
As shown in Fig. \ref{fig4}(b), $\lambda_{\rm max}$ reaches $1$ at
$T_c^{\rm DW}=73$meV for $x=0$ ($p_{\rm eff}=0.14$), while the spin-ordered
state is absent because $\alpha_s<1$.

%In the present 3D calculation, we employ $N=48^3$ $\k$
%meshes and $512$ Matsubara frequencies. In this case, the calculation
%accuracy is ensured for $T\geq 60$meV, which is higher than the
%experimental $T_c^{\rm RIXS}$. However, $T_c^{\rm DW}\ll60$meV can be
%obtained correctly by extrapolating the accurate results for $T\geq 60$meV.

As discussed in SM A \cite{SM}, we confirm that the obtained
results for $T\geq 60$meV are reliable.
Figure \ref{fig4}(c) shows the obtained $T$ dependence of $\lambda_{\rm max}$ for each
$x$ for $T\geq60$meV.
  We obtain almost perfect $T$-linear
  $\lambda_{\rm max}$ for all $x$, and it reaches unity for $x<0.05$.

  The $T$-linear $\lambda_{\q=\bm{0}}$ is generally realized in the
  paramagnon-interference mechanism, like Fe-based superconductors
  \cite{Tazai-LW}, which is consistent with the Curie-Weiss behavior of
  nematic susceptibility observed in experiments \cite{nem1,nem2}.
  Since the mechanism of the present 3D CDW is the same paramagnon interference, the
  $T$-linear behavior of $\lambda_{\rm max}$ is expected. \color{black}
By extrapolating the $T$-linear $\lambda_{\rm max}$ shown by dotted
line, we obtain reliable $T_c^{\rm DW}(<60$meV) for each $x\geq0.05$ shown in 
Fig. \ref{fig4}(d). 
This $x$ dependence of long-range order $T_c^{\rm DW}$ is consistent with
recent RIXS measurements \cite{Ni-CDW1,Ni-CDW2,Ni-CDW3,Ni-CDW4,Cu-CDW,Cu-CDW2}. 

At the bond-order QCP $(T_c^{\rm DW}=0)$, 
the non-Fermi-liquid transport phenomena and the strong pairing interaction
are induced by the bond-order fluctuations \cite{Tazai-kagome-CDW}.
In fact, non-Fermi-liquid transport
 phenomena \cite{Ni-Non-Fermi1,Ni-Non-Fermi2} and the enhancement of
 $T_{\rm SC}$ and $H_{c2}$ have been observed near the bond-order QCP.

In the present study, the obtained $q^c_z$ of the 3D CDW is
incommensurate.
In contrast, $q_z=0$ of the 3D CDW has been observed in YBCO
cuprates under a large magnetic field and a uniaxial strain
\cite{Cu-CDW-long,Cu-CDW-long2,Cu-CDW-long3,Cu-CDW-long4,Cu-CDW-long5}.
This significant difference is understood by the presence or absence of
the 3D secondary nesting.
We note that the RIXS peak $q_z\sim\pi$ in the 2D CDW phase
in YBCO is much broader along the $q_z$-direction than that of NdNiO$_2$
\cite{Ni-CDW3,Cu-CDW-long3,Cu-CDW-long4,Cu-CDW-long5}. This fact
means the realization of the 3D CDW in NdNiO$_2$.

%%%%%%%%%%%%%%%%%%%%%%%%%%%%%%%%%
%%%%%%%%%%%%%%%%%%%%%%%%%%%%%%%%%

Hereafter, we discuss differences in the CDW quantum critical behaviors between
$R$NiO$_2$ and cuprates.
The CDW in cuprates has been identified as the bond order by
the paramagnon-interference mechanism \cite{Cu-CDW-VC1,Cu-CDW-VC2,Cu-CDW-VC3}.
As shown in Fig. S1(c), the CDW instability at a fixed
$p_{\rm eff}$ in $R$NiO$_2$ is
stronger than that in cuprates. 
These differences are understood by the strength of the Coulomb
interaction: $U_1=3.8$eV in $R$NiO$_2$ given by the first-principles calculation is
larger than $U_1=2.6$eV in Hg cuprates \cite{Ni-Sakakibara}. 
In addition, the
3D bond order in
$R$NiO$_2$ is stabilized by the 
3D nesting $\q^c$ around R point shown in Fig. \ref{fig1}(e), which is absent in cuprates. For these reasons, the
CDW (bond-order) instability in $R$NiO$_2$ is stronger than that in cuprates.

 Note that the obtained
 long-range-order $T_c^{\rm DW}$ may be overestimated when the self-energy
 is not taken into account. In FeSe, the nematic transition temperature $\sim
 100$K is well reproduced by introducing the self-energy
 \cite{Tazai-LW}.
Noteworthy, this self-energy suppression is also important in the context of superconducting (SC) fluctuation
paraconductivity as discussed in  Ref. \cite{Dorin}.
Interestingly, the functional RG study \cite{Tsuchiizu-Ru} revealed that the AL diagrams of SC fluctuations give rise to the orbital order.
It is a fruitful future issue to include the SC fluctuations in the kernel function of the DW equation.

In recent years, strong correlation theories such as the DMFT \cite{DMFT1,DMFT2,DMFT3,DMFT4,DMFT5,DMFT6}, the
cluster DMFT \cite{CDMFT,CDMFT2}, the 2D density-matrix RG (DMRG) \cite{DMRG,DMRG2,DMRG3}, and the quantum Monte
Carlo methods \cite{CDMFT2,DiagQMC,DQMC} have made remarkable progress. Mott
insulators and
pseudogaps can be explained. On the other hand, the DW equation method, which
belongs to the weak correlation theory, is suitable for the analysis
of ``metallic ordered states'' such as CDW order \cite{Adv-review}. This
theory is applicable to various metallic systems, such as
``multi-orbital and multi-site models'' like iron-based superconductors
\cite{Onari-SCVC,Onari-form,Yamakawa-PRX,Onari-B2g,Onari-smectic,Onari-Frontier,Adv-review},
and kagome metals \cite{Tazai-kagome-CDW}. Notably, both the DW equation
method \cite{Kontani-sLC} and the DMRG study \cite{DMRG-sLC} lead to similar spin current orders in the square lattice Hubbard model. Thus, the DW equation theory is useful and reliable, and it is complementary to the above strong correlation theories.
\color{black}
 
% The cluster DMFT, the density matrix
% renormalization group, and diagrammatic Monte Carlo theories  can take the non-local quantum interference
% effects into account. Therefore, they are
% complementary to the present approach. \color{black}
 
In summary, 
we studied the origin of the 3D CDW in $R$NiO$_2$ based on a
realistic 3D Hubbard model.
We found that the 3D CDW is
identified as the $d$-wave bond order with the wave vector
$\q^c\sim(2\pi/3,0,q^c_z)$ $(0.2\pi\lesssim q^c_z\lesssim
2\pi/3)$. 
This 3D bond order with the period-three in the 
$x$-direction as shown in Fig. \ref{fig3}(c) \color{black} is driven by the
paramagnon-interference mechanism, and it is further stabilized
by the secondary 3D short-wavelength
nesting shown in Figs. \ref{fig1}(e) and (f).
The doping dependences of the CDW order have been well reproduced by the
present mechanism.
 The present paramagnon-interference mechanism would be a key concept toward a unified understanding of the CDW orders in nickelates and cuprates. This is an important issue for the future. \color{black}

%%%%%%%%%%%%%%%%%%%%%
\acknowledgements
We are grateful to 
%M. Koshino and
 Y. Yamakawa
for valuable discussions.
This work was supported
by Grants-in-Aid for Scientific Research from MEXT,
Japan (No. JP23H03299, No. JP19H05825, No. JP18H01175, and No. JP17K05543)

%%%%%%%%%%%%%%%%%%%%%%%
%\appendix
%\section{Supplemental Material}

%%%%%%%%%%%%%%%%%%%%%%%%
%references
%%%%%%%%%%%%%%%%%%%%%%%%

%\end{document}
%%%%%%%%%%%%%%%%%%%%%%%%%%%%%%%%%%%%%%%
\clearpage

\makeatletter
\renewcommand{\thefigure}{S\arabic{figure}}
\renewcommand{\theequation}{S\arabic{equation}}
\makeatother
\setcounter{figure}{0}
\setcounter{equation}{0}
\setcounter{page}{1}
\setcounter{section}{1}

\begin{widetext}
\begin{center}
{\bf 
[Supplementary Material] \\
Three-dimensional bond-order instability in infinite-layer
nickelates due to nonlocal quantum interference
}%
\end{center}

\begin{center}
Seiichiro Onari and Hiroshi Kontani
\end{center}

\begin{center}
\textit{Department of Physics, Nagoya University, Nagoya 464-8602, Japan}
\end{center}

\end{widetext}
\subsection{A: Model Hamiltonian of NdNiO$_2$, formalism of the RPA and the DW equation}
\begin{figure}[!htb]
\includegraphics[width=.99\linewidth]{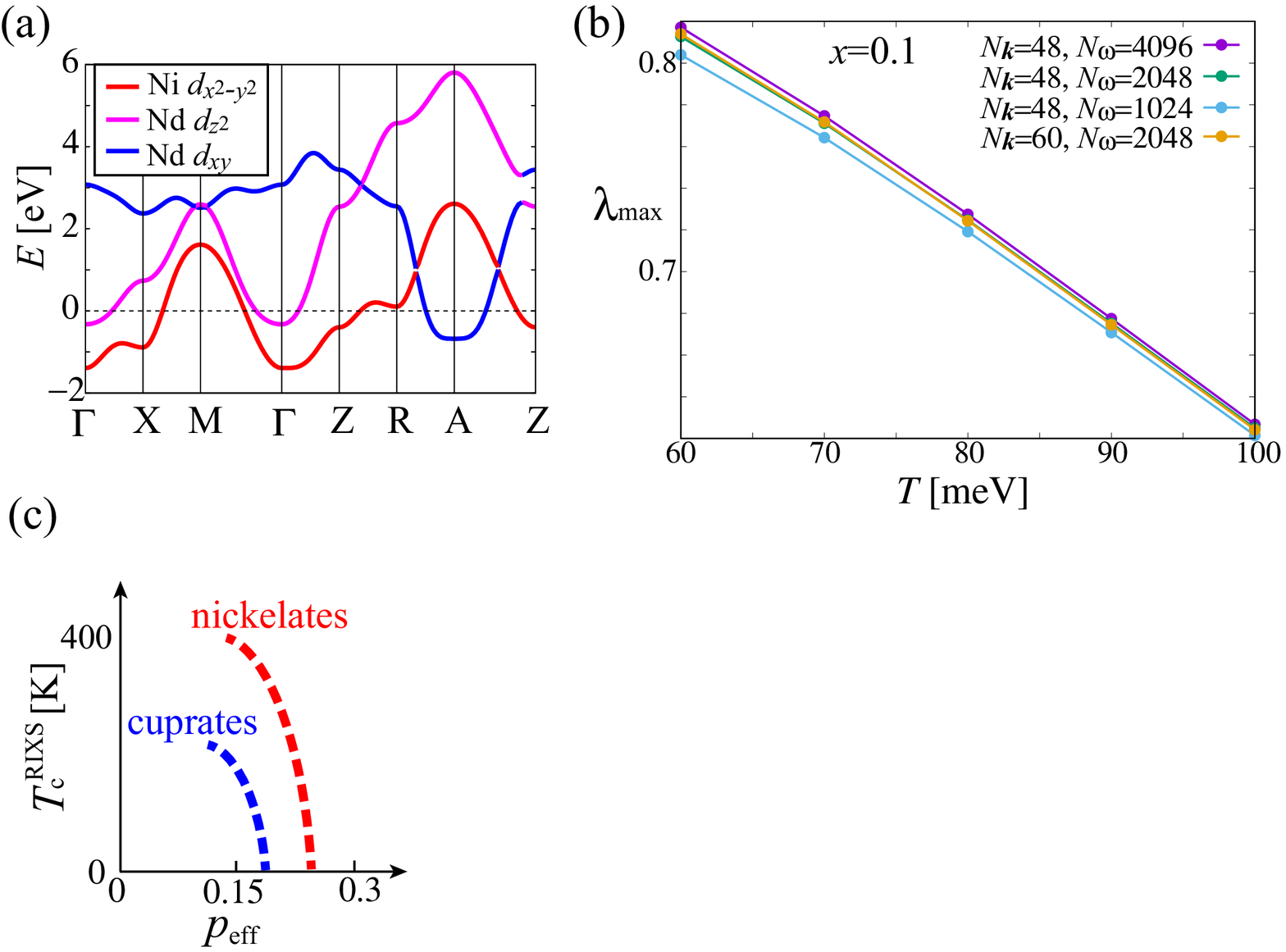}
 \caption{
 (a) Band dispersion of NdNiO$_2$.
 (b) $T$ dependences of $\lambda_{\rm max}$ at $x=0.1$
 for several $N_{\bm{k}}$ and $N_\omega$.
 (c) Schematic $p_{\rm eff}$
 dependences of $T_c^{\rm RIXS}$ in
 nickelates and overdoped cuprates.
 }
\label{figS0}
\end{figure}

First, we introduce a tight-binding model for NdNiO$_2$ by referring 
the tight-binding model in Ref. \cite{S-Ni-first-principle4}.
We modify the hoppings of Ni $d_{x^2-y^2}$ orbital in order to reproduce
the parallel FSs around the M point, which have been obtained in other
first-principles calculations
\cite{S-Ni-Sakakibara,S-Ni-first-principle5}.
We modify the intralayer next-nearest-neighbor
hopping $t_2$, the intralayer third-nearest one $t_3$, 
interlayer nearest one $t_z$, and next-nearest-interlayer one $t_{z2}$ to
$t_2/t_1=-1/3$, $t_3/t_1=0.2$, $t_z/t_1=2/3$, and  $t_{z2}/t_1=-0.165$ ($t_1$ is the nearest-neighbor
hopping), respectively. These intralayer hoppings are similar to those
in YBCO cuprate. The obtained band dispersion is shown in Fig. \ref{figS0}(a).

Here, we explain the Coulomb interaction introduced in the present
study.
Ni $d_{x^2-y^2}$ orbital, Nd $d_{z^2}$ orbital, and Nd $d_{xy}$ orbital
are included in our model. We introduced the on-site Coulomb interaction
$U_1$ for Ni $d_{x^2-y^2}$ orbital. 
By using the Coulomb interaction,
the spin (charge) susceptibility for orbital 1 in the RPA is given by  
\begin{equation}
{\chi}^{s(c)}(q)={\chi^0}(q)[1-(+)U_1{\chi^0(q)}]^{-1},
\end{equation}
where the irreducible susceptibility is
\begin{equation}
\chi^0(q)= -\frac{T}{N}\sum_k
G(k+q)G(k).
\end{equation}
$G(k)$ is the Green function for orbital 1 without self-energy 
${G}(k)=[\frac{1}{(i\e_n-\mu){\hat1}-{\hat{h}}^0(\k)}]_{1,1}$ 
for $=[\k,\e_n=(2n+1)\pi T]$. 
Here, ${\hat{h}}^0(\k)$ is the matrix expression of $H^0$ 
and $\mu$ is the chemical potential.

%
%Figure \ref{figS0} shows $q_x$ dependence of
%$\chi^0_{1,1;1,1}(q_x,0,q_z)$ for each $q_z$.
%$\chi^0_{1,1;1,1}(q_x,0,q_z)$ has a broad peak at $\sim (2\pi/3,0,q_z)$
%for $q_z<2\pi/3$, which is induced by the secondary nesting between
%the parallel FSs around the M point in
%Fig. \ref{fig1}(e).  

%%%%%%%%%%%%%%%%%%%%%%%%%%%%%%%%%%
%%%%%%%%%%%%%%%%%%%%%%%%%%%%%%%%%%

In the present study, we set $U_1=1.43$eV.
We use $N=N_{\bm{k}}^3=48\times 48 \times 48$ $\k$ meshes and $N_\omega=4096$ Matsubara frequencies.

The four-point vertex ${I}^{\bm{q}}(k,k')$ for orbital 1 in the DW
Eq. (\ref{eqn:linearized}) is given as
\begin{eqnarray}
&& \!\!\!\!\!\!\!\!\!\!\!
I^{\bm{q}}(k,k')=\sum_{b=s,c}
\left[-\frac{a^b}{2} V^{b}(k-k')\right.
\nonumber \\
&& 
+\frac{T}{N}\sum_{p}
 \frac{a^b}{2} V^{b}\left(p+\frac{\q}{2}\right)V^{b}\left(p-\frac{\q}{2}\right)
 \nonumber \\
&& \qquad\qquad
\times G(k-p)G(k'-p)
\nonumber \\
&&
+\frac{T}{N}\sum_{p}
 \frac{a^b}{2} V^{b}\left(p+\frac{\q}{2}\right)V^{b}\left(p-\frac{\q}{2}\right)
 \nonumber \\
&& \qquad\qquad
\left.\times G(k-p)G(k'+p)\right],
%-({\rm Double\;counting\;} [\hat{\Gamma}^{s(c)}]^2 \;{\rm terms})
\label{eqn:S-K} 
\end{eqnarray}
%\begin{eqnarray}
%&& \!\!\!\!\!\!\!\!\!\!\!
%K^{\bm{q}}_{l,l';m,m'}(k,k')=
%\sum_{l'',m''}\left[\frac32 V^{s}(k-k')+\frac12 V^{c}(k-k')\right]_{l,l'';l',m''}
%\nonumber \\
%&& \ \ \ \ \ \ \ \ \ \ 
%\times G^0_{m',m''}\left(k'-\frac{\q}{2}\right)G^0_{l'',m}\left(k'+\frac{\q}{2}\right)
% \nonumber \\
%&&
%-\frac{T}{N}\!\!\!\!\sum_{p,l_1,l_2,l_3,m_1,m_2,m_3}\!\!\!\!\!\!\!\!\!\!\!\!\!\!
% \left[ \frac32 V^{s}_{l_1,l_2;l',m_1}\left(p-\frac{\q}{2}\right)V^{s}_{l,l_3;m_2,m_3}\left(p+\frac{\q}{2}\right) \right.
% \nonumber \\
%&& 
%\left. +\frac12 V^{c}_{l_1,l_2;l',m_1}\left(p-\frac{\q}{2}\right)V^{c}_{l,l_3;m_2,m_3}\left(p+\frac{\q}{2}\right) \right]G^0_{l_3,m_1}(k-p)
%\nonumber \\
%&& 
%\times \left[{\Lambda}^\q_{m',l_1;l_2,m_3;m_2,m}(k';p)+{\Lambda}^\q_{m',m_3;m_2,l_1;l_2,m}(k';-p)\right],
%%-({\rm Double\;counting\;} [\hat{\Gamma}^{s(c)}]^2 \;{\rm terms})
%\label{eqn:S-K} 
%\end{eqnarray}
%
where $a^s=3$, $a^c=1$, $p=(\p,\w_l)$, and
${V}^{s(c)}(q)=+(-)U_1+U_1^2{\chi}^{s(c)}(q)$. 
$\Q$
and $\Q'$ in Fig. \ref{fig1} (a) are given by $\Q=\bm{p}+\q/2$ and
$\Q'=\bm{p}-\q/2$. \color{black}
%in Ref. \cite{Onari-SCVC,Onari-SCVCS}.

In Eq. (\ref{eqn:S-K}),
the first line corresponds to the MT term,
and the second and third lines give the AL1 and AL2 terms, respectively.
In the MT term,
the first-order term with respect to $U_1$ 
gives the Hartree--Fock term in the mean-field theory.

Here, the number of Matsubara frequencies
$N_\omega=4096$ is sufficient for $T\geq 60$meV because of the relation $2\pi T
N_\omega \gtrsim 100W_{\rm band}\sim 400$eV.
However, due to the
lack of $\k$ meshes $N=N_{\bm{k}}^3=48\times 48 \times 48$, the obtained
$\lambda_{\rm max}$ would be underestimated at low temperatures. To
verify the numerical accuracy
for $T\geq 60$meV, we show the $T$ dependence of $\lambda_{\rm max}$ at
$x=0.1$ for each $N_{\bm{k}}=48\sim60$ and $N_\omega=1024\sim4096$ in Fig. \ref{figS0}(b).

Figure \ref{figS0}(c) shows schematic $p_{\rm eff}$
 dependences of $T_c^{\rm RIXS}$ observed
 in nickelates and overdoped cuprates. 
 $T_c^{\rm RIXS}\lesssim 400$K in nickelates is higher than $T_c^{\rm RIXS}\lesssim 200$K in cuprates. \color{black}

\subsection{B: Detailed results of the form factor}

\begin{figure}[!htb]
\includegraphics[width=.99\linewidth]{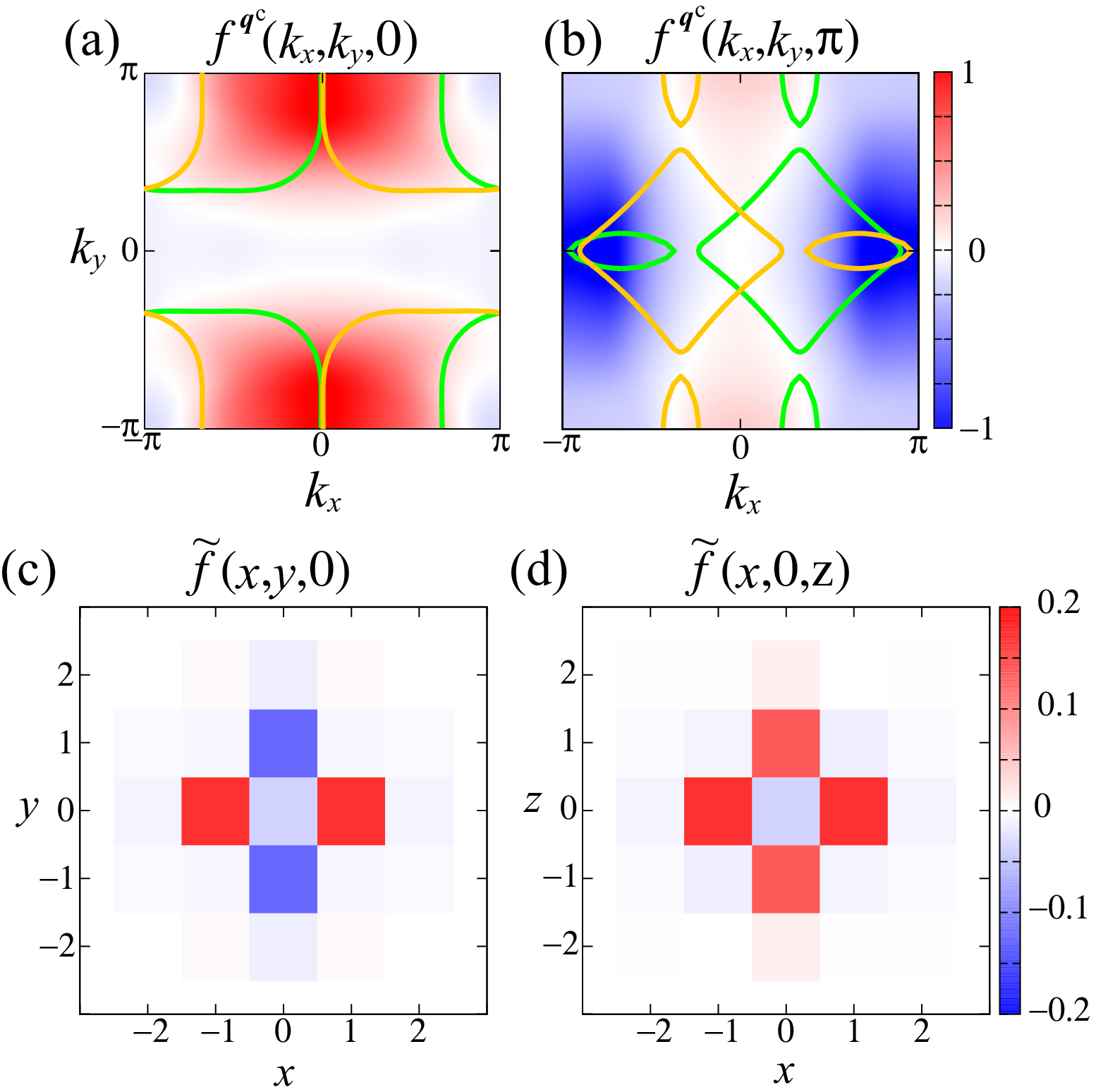}
 \caption{
(a) $\k$ dependence of $f^{\q^c}(\k)$ for $\q^c=(2\pi/3,0,0.4\pi)$ at
 $k_z=0$ plane, and (b) that at $k_z=\pi$ plane, where green and orange
 lines denote FSs shifted by $\q^c/2$ and $-\q^c/2$, respectively.
 (c) $\bm{r}$ dependence of $\tilde{f}(\bm{r})$ at $z=0$ plane, and (d) that at
 $y=0$ plane.
 }
\label{figS1}
\end{figure}

Here, we explain details of the obtained $\q^c=(2\pi/3,0,0.4\pi)$ form factor.
 Figure \ref{figS1}(a) and (b) show the form factors
 $f^{\q^c}(\k)$ for orbital 1, which is derived from
 $f^{\q^c}(k)$.  $f^{\q^c}(\k)$ is obtained by simple
 extrapolation of the lowest Matsubara frequency component and the
 second lowest one. The obtained $f^{\q^c}(\k)$ is not sensitive to the method of analytical connection, which is verified by the results of the Pade approximation. \color{black}
 FSs shifted by $\pm\frac{\q^c}{2}$
 are shown by green and orange lines.
The form factor has a large value at $\k$, where the two shifted FSs
 overlap since the $g^{\q^c}(k')$ in the kernel function is enlarged
 there. The sign change of $f^{\q^c}(k)$ is induced by the MT term,
 $f^{\q^c}(\k)f^{\q^c}(\k-\Q_s)<0$ $[\k\sim (2\pi/3,0,\pi)$
 ,$\Q_s\sim(2\pi/3,2\pi/3,\pi)]$. From the $\k$ dependence of the form factor, we find that the
 form factor corresponds to the $d_{x^2-y^2}$-wave bond order.

 In order to study the real space structure of the bond order, we calculate a Fourier
transformed form factor $\tilde{f}(\bm{r})$ given as
 \begin{equation}
\tilde{f}(\bm{r})=\sum_{\bm{k}}f^{\q^c}(\bm{k})e^{-i\bm{k}\cdot\bm{r}}.  
 \end{equation}
The obtained $\tilde{f}(\bm{r})$ is shown in Figs.
\ref{figS1}(c) and (d). The values at the nearest neighbors from $\bm{r}=\bm{0}$
are dominant. Sign of $\tilde{f}(\pm1,0,0)$ is opposite to that of
$\tilde{f}(0,\pm1,0)$, which means the $d_{x^2-y^2}$-wave bond order in
$xy$ plane. Since the magnitude of $\tilde{f}(0,0,\pm1)$ along $z$-direction is similar
to that of $\tilde{f}(\pm1,0,0)$, the bond order has
3D structure.
In addition, the small value of $\tilde{f}(\bm{0})\ne0$ means
emergence of the slight charge order.

The modulation of hopping $\delta t_{i,j}$ is given by
\begin{equation}
 \delta t_{i,j}=2\tilde{f}(\bm{r_i}-\bm{r_j})\cos\left[\frac{\bm{q}^c}{2}\cdot(\bm{r_i}+\bm{r_j})\right].
\end{equation}
The obtained period-three $\delta t_{i,j}$ bond order around $z=0$ plane is shown in
Fig. \ref{fig3}(c). Since $\q^c$ has $q_z$ component
$0.2\pi\lesssim q^c_z\lesssim 2\pi/3$, the bond order also modulates along the $z$-direction.

%%%%%%%%%%%%%%%%%%%%%%%%%%%%%%%%%%
%%%%%%%%%%%%%%%%%%%%%%%%%%%%%%%%%%

%%%%%%%%%%%%%%%%%%%%%%%%
%references
%%%%%%%%%%%%%%%%%%%%%%%%

\end{document}